\documentclass[rmp,11pt,tightenlines,letterpaper,twoside,nobibnotes,altaffilletter,amssymb,showpacs,byrevtex]{revtex4}
\usepackage{latexsym}
\usepackage{times}
\bibpunct{}{}{,}{s}{}{\textsuperscript{,}}

\begin{document}
\bibliographystyle{apsrev}

\pagestyle{myheadings}
\markboth{Irrep's of the extended Poincar\'e group}{R. O. de Mello and V. O. Rivelles}

\title{The irreducible unitary representations of the extended Poincar\'e group
in (1+1) dimensions}\thanks{Revised version of paper published in J. Math. Phys. \textbf{45}, 1156 (2004).}
\author{R. O. \surname{de Mello}}\email[Electronic mail: ]{romello@ift.unesp.br}
\affiliation{Instituto de F\'\i sica Te\'orica, Universidade Estadual Paulista\\
R.Pamplona 145, 01405-900, S.Paulo, SP, Brazil}
\author{V. O. \surname{Rivelles}}\email[Electronic mail: ]{rivelles@fma.if.usp.br}
\affiliation{Instituto de F\'\i sica, Universidade de S\~ao Paulo\\
C.Postal 66318, 05315-970, S.Paulo, SP, Brazil}

\date{\today}

\begin{abstract}
We prove that the extended Poincar\'e group in (1+1) dimensions $\bar{\mathcal{P}}$
is non-nilpotent solvable exponential, and therefore that it belongs to type I. We determine its first and
second
cohomology groups in order to work out a classification of the two-dimensional
relativistic elementary systems. Moreover, all 
irreducible unitary representations of $\bar{\mathcal{P}}$ are
constructed by the orbit method. 
The most physically interesting class of irreducible
representations corresponds to the anomaly-free relativistic particle in (1+1) dimensions,
which cannot be fully quantized. However, we show that the corresponding coadjoint orbit of
$\bar{\mathcal{P}}$ determines a covariant maximal polynomial quantization by unbounded
operators, which is enough to
ensure that the associated quantum dynamical problem can be consistently solved, thus
providing a physical interpretation for this particular class of representations.  
\end{abstract}


\maketitle

\section{\label{intro}Introduction}
Much of the interest in the extended Poincar\'e group in (1+1)
dimensions $\bar{\mathcal{P}}$ stems from the fact
that the Callan--Giddings--Harvey--Strominger (CGHS) model of two-dimensional dilatonic gravity \cite{cghs}  may be formulated as a gauge theory \cite{jackiw2} of $\bar{\mathcal{P}}$. The
``string-inspired'' CGHS theory is particularly interesting because it generates an exactly
solvable model of quantum gravity, which allows the investigation of several aspects of quantum
black hole physics. \cite{kazama, strominger} An outstanding problem in this context is the
coupling of matter sources in an extended Poincar\'e
gauge-invariant fashion. \cite{jackiw, lupi, rivelles}  

The main purpose of this paper is to prove that $\bar{\mathcal{P}}$ is solvable exponential, so that the Bernat--Pukanszky theory
of exponential groups \cite{bernat, puk} can be strictly applied to work out all its unitary irreducible representations (irrep's). 
Some of these irrep's
were presented in Gadella \textit{et al.} \cite{gadella} but, although it was mentioned \cite{negro}
that these irrep's were calculated by the Mackey theory
and the orbit method, it was not shown that $\bar{\mathcal{P}}$ has a regular semidirect product
structure, nor that $\bar{\mathcal{P}}$ is solvable exponential.
These authors adopt the same point of view as
that of Cari\~nena \textit{et al.}, \cite{carinena} which should be contrasted with ours.

Our approach to the two-dimensional relativistic elementary systems
is similar to that which was adopted by Azc\'arraga and Izquierdo \cite{azcarraga}
with respect to a nonrelativistic particle of unit charge in a constant magnetic field.
Indeed, we show
in this paper that the coadjoint orbit corresponding to the anomaly-free relativistic particle in (1+1) dimensions determines a covariant maximal
polynomial quantization for it, which provides a physical interpretation for the associated
class of irrep's of $\bar{\mathcal{P}}$.

This paper is organized as follows. In Sec.~\ref{sec:1}, we show that $\bar{\mathcal{P}}$ is solvable
exponential and calculate its first and second cohomology groups. In
Sec.~\ref{sec:4}, we determine the coadjoint orbits of $\bar{\mathcal{P}}$ in order to classify the two-dimensional relativistic elementary systems and to work out
explicitly all the irrep's of $\bar{\mathcal{P}}$.
In Sec.~\ref{sec:5}, we provide a physical interpretation for a particular class of irrep's of $\bar{\mathcal{P}}$ through a covariant maximal polynomial quantization of the
anomaly-free
relativistic particle in (1+1)
dimensions. Finally, in
Sec.~\ref{sec:6} we draw our conclusions and discuss further possible developments.
We sketch the method of orbits in the Appendix, in order to provide some supplementary
material for the understanding of Sec.~\ref{sec:4}.

\section{\label{sec:1}The Extended Poincar\'e Group in (1+1)
dimensions $\bar{\mathcal{P}}$}

The extended Poincar\'e algebra $\bar{\textrm{\i}}^{1}_{2}$ is defined by means of an
unconventional contraction of a pseudoextension \cite{azcarraga} of the anti-de Sitter algebra
so(2,1) as
\begin{equation}\label{eq:algpoincest}
\lbrack P_{a},J\rbrack = \sqrt{-h}\varepsilon^{\verb+ +b}_{a}P_{b}\textrm{,}\quad
\lbrack P_{a},P_{b}\rbrack=B\varepsilon_{ab} I\textrm{,}\quad\textrm{and}\quad
\lbrack P_{a},I\rbrack  =  \lbrack J,I\rbrack=0\textrm{,}
\end{equation}
where $a,b\in\{0,1\}$, $\varepsilon^{01}=-\varepsilon_{01}=1$, and the indices $a$ and $b$ are
raised and lowered by the metric $h_{ab}=\textrm{diag}(1,-1)$ with $h:=\textrm{det}\,h_{ab}=-1$. Throughout this paper, we shall adopt
units where $c=1$.
The generators of
translations are $P_{a}:=\bar{T}_{a}$, and their dimensions are $L^{-1}$. The generator of
Lorentz transformations is $J:=\bar{T}_{2}$, which is dimensionless. The central generator, the
dimension of which is $[\hbar]^{-1}$, is $I:=\bar{T}_{3}$, and the central charge has
dimension $[B]=L^{-2}\times[\hbar]$. 

The group law $g''(\theta''^{a},\alpha'',\beta'')=g'(\theta'^{a},\alpha',\beta')g(\theta^{a},\alpha,\beta)$ determined by Eq.~(\ref{eq:algpoincest}) is given by
\begin{equation}\label{eq:leicompparam}
\theta''^{b} = \theta'^{b}+\Lambda(\alpha')^{b}_{\verb+ +a}\theta^{a},\quad
\alpha'' = \alpha'+\alpha,\quad\textrm{and}\quad
\beta'' = \beta'+\beta+\frac{B}{2}\theta'^{c}\varepsilon_{cb}\Lambda(\alpha')^{b}_{\verb+ +a}
\theta^{a},
\end{equation}
where $\Lambda(\alpha)^{a}_{\verb+ +b}=\delta^{a}_{\verb+ +b}\cosh\alpha+\sqrt{-h}\,\varepsilon^{a}_{\verb+ +b}\sinh\alpha$,
and it corresponds to the coset decomposition
$g(\theta^{a},\alpha,\beta)=\exp(\theta^{a}P_{a})\exp(\alpha J)
\exp(\beta I)$.
The adjoint representation of $\bar{\mathcal{P}}$ is given by
\begin{displaymath}
(\textrm{Ad}\,g)^{A}_{\verb+ +B}=
\left(\begin{array}{ccc}
\Lambda^{a}_{\verb+ +b} & \theta^{c}\varepsilon_{c}^{\verb+ +a}\sqrt{-h} & 0\\
0 & 1 & 0\\
B\theta^{c}\varepsilon_{cd}\Lambda^{d}_{\verb+ +b} & -\frac{B}{2\sqrt{-h}}\theta^{a}\theta_{a} & 1
\end{array}\right),
\end{displaymath}
and the invariant Casimir operator determines the metric $h_{AB}$ such that
$\langle V,V\rangle=h^{AB}V_{A}V_{B}=V^{a}V_{a}-2(B/\sqrt{-h})V_{2}V_{3}$,
for any vector $V=V^{A}\bar{T}_{A}$ in $\bar{\textrm{\i}}^{1}_{2}$, with $A,B\in\{0,1,2,3\}$. 
The dimensions of the metric components are
$[h_{ab}]=L^{-2}$, and $[h_{23}]=[h_{32}]=[\hbar]^{-1}$.

The extended Poincar\'e algebra has the structure of a semidirect product
$\bar{\textrm{\i}}^{1}_{2} = \textrm{so}(1,1)\times_{\rho} \textrm{wh}$, where
$\textrm{so}(1,1)=\Re$ is the Abelian
subalgebra generated by $J$, and
wh is the maximal nilpotent ideal spanned by
$\{P_{0},P_{1},I\}$, which is isomorphic to the Lie
algebra of the Weyl--Heisenberg group WH. The representation $\rho$ of $\textrm{so}(1,1)$ on
wh is given by the restriction of the adjoint representation of $\bar{\textrm{\i}}^{1}_{2}$ to
$\textrm{so}(1,1)$.

It is well-known that $\bar{\textrm{\i}}^{1}_{2}$ is solvable; \cite{jackiw}
however, it is also not nilpotent, since its descending central series,
$\bar{\textrm{\i}}^{1^{1}}_{2}=\bar{\textrm{\i}}^{1}_{2}$, $\bar{\textrm{\i}}^{1^{2}}_{2}=
[\bar{\textrm{\i}}^{1}_{2},\bar{\textrm{\i}}^{1^{1}}_{2}]=\textrm{wh}$,\ldots,
$\bar{\textrm{\i}}^{1^{k}}_{2}=[\bar{\textrm{\i}}^{1}_{2},\bar{\textrm{\i}}^{1^{k-1}}_{2}]=
\textrm{wh}
\quad\forall k\ge 2$,
does not vanish for any value of $k$.
It is also not difficult to see that the extended Poincar\'e group
$\bar{\mathcal{P}}$ and its Lie algebra
$\bar{\textrm{\i}}^{1}_{2}$ are solvable exponential, since for any $X\in \bar{\textrm{\i}}^{1}_{2}$ the eigenvalues of
$\textrm{ad}(X)$ are all
real. \cite{conze}

As a consequence, $\bar{\mathcal{P}}$ is defined as the connected
and simply connected image of $\bar{\textrm{\i}}^{1}_{2}$ by the exponential mapping
$\bar{\mathcal{P}}=\exp(\bar{\textrm{\i}}^{1}_{2})$, and every
element $g\in\bar{\mathcal{P}}$ belongs to a one-parameter subgroup, so the group law given by
Eq.~(\ref{eq:leicompparam}) holds globally. Another consequence is that $\bar{\mathcal{P}}$ is homologically trivial; therefore, the Van Est theorem \cite{azcarraga} ensures that the cohomology groups on
$\bar{\mathcal{P}}$ are canonically isomorphic to the corresponding cohomology groups on
$\bar{\textrm{\i}}^{1}_{2}$.

The first cohomology group of $\bar{\textrm{\i}}^{1}_{2}$ can be readily calculated,
$H^{1}_{0}(\bar{\textrm{\i}}^{1}_{2},\Re)=(\bar{\textrm{\i}}^{1}_{2}/[\bar{\textrm{\i}}^{1}_{2},\bar{\textrm{\i}}^{1}_{2}])^{*}=\Re$. In order to work out the second cohomology group,
it is enough to show that the space of two cocycles $Z^{2}_{0}(\bar{\textrm{\i}}^{1}_{2},\Re)\subset \Lambda^{2}\bar{\textrm{\i}}^{1}_{2}$ has the same dimension of the space of two coboundaries $B^{2}_{0}(\bar{\textrm{\i}}^{1}_{2},\Re)$. It turns out that $\textrm{dim}\, H^{2}_{0}(\bar{\textrm{\i}}^{1}_{2},\Re)=\textrm{dim}\,
Z^{2}_{0}(\bar{\textrm{\i}}^{1}_{2},\Re)-\textrm{dim}\,
B^{2}_{0}(\bar{\textrm{\i}}^{1}_{2},\Re)=0$.

\section{\label{sec:4}Construction of the Irrep's of $\bar{\mathcal{P}}$ by
its Coadjoint Orbits}

It will be shown in Sec.~\ref{sec:5} that the anomaly-free 
Lagrangian describing a relativistic particle in flat two-dimensional space--time must be
invariant under $\bar{\mathcal{P}}$, consistently with $H^{2}_{0}(\bar{\mathcal{P}},\Re)=0$.
It follows that the relevant dynamical group in two dimensions is $\bar{\mathcal{P}}$, so the adequate statement of the principle of relativity
in (1+1) dimensions should require that the equations of motion are covariant under the transformations
of $\bar{\mathcal{P}}$.

This means that the elementary particles in (1+1) dimensions must belong to irrep's of $\bar{\mathcal{P}}$ at the quantum level, and constitute relativistic
elementary systems in this sense. On the other
hand, the group-theoretic approach is concerned about a corresponding
notion of elementary system at the classical level, i.e., a system that cannot be decomposed
into smaller parts without breaking the symmetry. \cite{kirillov4}
It turns out
that the  irreducibility condition  is translated naturally into a transitivity one at the
classical level; therefore,
a classical elementary system is defined as a homogeneous symplectic manifold (HSM). We say
that 
an elementary system $(S,\Omega)$ is a Hamiltonian G-space, \cite{kostant} or a strictly
homogeneous symplectic manifold, if further the dynamical group $G$
possesses a Poisson action upon $S$.       

We recall that, due to the Kirillov theorem, \cite{kirillov} every HSM associated with some dynamical group $G$ is locally isomorphic to a coadjoint orbit of
$G$ or to a coadjoint orbit of the central extension of $G$ by $\Re$. Then, if further all the
coadjoint orbits of $G$ are simply connected and
$H^{2}_{0}(\mathfrak{g},\Re)=0$, then the momentum mapping will be a
symplectomorphism between every classical
elementary system $(S,\Omega)$ upon which the action of $\mathfrak{g}$ is globally Hamiltonian
and a certain coadjoint orbit.

Applying this theorem, we discover that every classical relativistic elementary system upon
which the action of $\bar{\textrm{\i}}^{1}_{2}$ is
globally Hamiltonian is simply connected, and symplectomorphic to one of the
coadjoint orbits of $\bar{\mathcal{P}}$ that are calculated below, since it is a connected solvable
exponential Lie
group with $H^{2}_{0}(\bar{\textrm{\i}}^{1}_{2},\Re)=0$ (see Sec.~\ref{sec:1}).
Although this classification does not exhaust all the two-dimensional relativistic elementary systems, since
$H^{1}_{0}(\bar{\mathcal{P}},\Re)=\Re$, it is general enough to include the most
physically interesting cases, such as the anomaly-free relativistic particle in (1+1) dimensions.

The coadjoint orbit through
$\zeta=\zeta_{A}\bar{\omega}^{A}$ in $\bar{\textrm{\i}}^{1^{*}}_{2}$
is formed by the points
$\mu=u_{A}\bar{\omega}^{A}$ satisfying $u_{A}=\zeta_{B}(\textrm{Ad}\,g^{-1})^{B}_{\verb+ +A}$, 
where $\{\bar{\omega}^{A}\}$ is the basis of
$\bar{\textrm{\i}}^{1^{*}}_{2}$ dual to $\{\bar{T}_{A}\}$.
As a consequence, the following identities hold: $u^{A}u_{A}=\zeta^{A}\zeta_{A}$ and
$u_{3}=\zeta_{3}$. The stability group of $\zeta\in\bar{\textrm{\i}}^{1^{*}}_{2}$
is generated by the subalgebra
$\bar{\textrm{\i}}^{1}_{2\zeta}\subset \bar{\textrm{\i}}^{1}_{2}$, which is the kernel of the
Kirillov two form $B_{\zeta}(X,Y)$, formed by the vectors $Y\in\bar{\textrm{\i}}^{1}_{2}$ for
which $\langle\zeta,[X,Y]\rangle=0$, $\forall X\in\bar{\textrm{\i}}^{1}_{2}$. The dimension of the coadjoint orbit can be deduced from the dimension of the stability group.

As the space of coadjoint orbits of $\bar{\mathcal{P}}$ parametrizes both the set of
relativistic elementary systems in two dimensions and the unitary dual of $\bar{\mathcal{P}}$,
we will present the coadjoint orbits of $\bar{\mathcal{P}}$ together with their
associated irrep's. The problem splits into three cases, and we will follow the methodology
sketched in the Appendix for working out all the irrep's of $\bar{\mathcal{P}}$. 

Since $\textrm{ad}(X)$ is traceless for all $X\in\bar{\textrm{\i}}^{1}_{2}$,
$\bar{\mathcal{P}}$ is unimodular (i.e., $\Delta_{\bar{\mathcal{P}}}=1$). Also,
because the real eigenvalues of $\textrm{ad}(X)$ are not all zero for every
$X\in\bar{\textrm{\i}}^{1}_{2}$,
$\bar{\mathcal{P}}$ is not quasinilpotent (see the Appendix).
Consequently, in order to apply the method of
orbits to $\bar{\mathcal{P}}$,
we must find for any $\zeta\in\bar{\textrm{\i}}^{1^{*}}_{2}$ a subalgebra
$\mathfrak{h}\subset\bar{\textrm{\i}}^{1}_{2}$ of a maximal dimension in the family of
the subalgebras subordinate to $\zeta$, further satisfying Pukanszky's condition.

\subsection{Case $\zeta_{3}\ne 0$}
The
coadjoint orbit is the two-dimensional surface diffeomorphic to
$\Re^{2}$ in the three-dimensional hyperplane $u_{3}=\zeta_{3}$, defined by the equations
\begin{equation}\label{eq:coorbzeta3}
u_{2}  =  \frac{u^{a}u_{a}\sqrt{-h}}{2Bu_{3}}-\frac{\zeta^{A}\zeta_{A}\sqrt{-h}}{2Bu_{3}}
\quad\textrm{and}\quad
\quad u_{3}  = \zeta_{3},
\end{equation}
and passing through the point
$\zeta=(0,0,-[(\zeta^{A}\zeta_{A}\sqrt{-h})/2B\zeta_{3}],\zeta_{3})$. These coadjoint
orbits are classified by $\zeta_{3}$ and $\zeta^{A}\zeta_{A}$.

Since we may choose any point on the
coadjoint orbit (see the Appendix), we pick $\zeta$.
Denoting by $(J,P_{+},I)$ the subalgebra of $\bar{\textrm{\i}}^{1}_{2}$ spanned by these
vectors, where
$P_{+}=P_{0}+P_{1}$, it is clear that $\mathfrak{h}=(J,P_{+},I)$ is subordinate to $\zeta$,
since its first derived algebra is $[\mathfrak{h},\mathfrak{h}]=(P_{+})$, which is orthogonal
to $\zeta$ or $\langle\zeta,(P_{+})\rangle=0$.
The subalgebra $\mathfrak{h}$ subordinate to
$\zeta$ is also admissible, since its codimension is 1, which is half the dimension of the
coadjoint orbit, and it satisfies Pukanszky's condition
$\zeta + \mathfrak{h}^{\bot}\subset\textrm{orb}(\zeta)$. Since any other admissible subalgebra leads to a unitary
equivalent representation (see the Appendix), we choose $\mathfrak{h}$.

The typical element of the subgroup $H$ generated by $\mathfrak{h}$ will be denoted by $h(\theta^{+},\alpha,\beta)=\exp(\theta^{+}P_{+})\exp(\alpha J)\exp(\beta I)$,
so that we can  define (see the Appendix) the one-dimensional representation of $H$ by
$\chi(\theta^{+},\alpha,\beta)=U(h(\theta^{+},\alpha,\beta))=\exp\left(i(-\alpha[(
\zeta^{A}\zeta_{A}\sqrt{-h})/2B\zeta_{3}]+\beta\zeta_{3})\right)$.
The adjoint representation of the subgroup $H$ can be straightforwardly calculated, so that
the modulus of $H$ is given by $\Delta_{H}(h)= \vert \textrm{det}(\textrm{Ad}\,h)\vert^{-1}=e^{\alpha}$.  
The space $L(\bar{\mathcal{P}},H,U)$ invariant under right translations on
$\bar{\mathcal{P}}$ is formed by the complex functions satisfying the condition
(see the Appendix)
\begin{eqnarray}\label{eq:rightinvfunc}
F(h(\theta^{+'},\alpha',\beta')\cdot g(\theta^{a},\alpha,\beta))  =  
e^{-(\alpha'/2)}\chi(\theta^{+'},\alpha',\beta')F(g(\theta^{a},\alpha,\beta)),
& &\nonumber\\
F\left(g\left(\Lambda^{a}_{\verb+ +b}(\alpha')\theta^{b}+\theta^{+'},\alpha'+\alpha,
\beta'+\beta+\frac{B}{2}\theta^{+'}e^{\alpha'}(\theta^{0}-\theta^{1})\right)\right)  & &
\nonumber\\
=e^{-(\alpha'/2)}\exp\left(i\left(-\alpha'\frac{
\zeta^{A}\zeta_{A}\sqrt{-h}}{2B\zeta_{3}}+\beta'\zeta_{3}\right)\right)F(g(\theta^{a},\alpha,\beta)). & &
\end{eqnarray}
This means that the space $L(\bar{\mathcal{P}},H,U)$ is determined by the value of $F$ at
$\theta^{0}=\alpha=\beta=0$.  

It is not difficult to see that every element of
$\bar{\mathcal{P}}$ can be uniquely written as $g=h\cdot k$, where $h\in H$, $k\in K$, and $K$ is
the one-parameter subgroup of $\bar{\mathcal{P}}$ generated by $P_{1}\in
\bar{\textrm{\i}}^{1}_{2}$. 
Choosing the Borel mapping $s(x):=k$, where $x\in X=H\backslash\bar{\mathcal{P}}$ and
$x=Hg=Hhk=Hk$, we can identify the right-coset space $X$ with the subgroup
$K\subset\bar{\mathcal{P}}$, in the sense that $s(X)=K$. The bi-invariant measure on
$\bar{\mathcal{P}}$ splits into $d\mu(g)=\Delta_{H,\bar{\mathcal{P}}}(h)d\nu_{s}(x)d\nu(h)$,
where the measure on $X$ is determined by the right Haar measure on $K=\Re$,
$d\nu_{s}(x)=d\nu(s(x))$, which is only $\bar{\mathcal{P}}$-quasi invariant, because
$\Delta_{\bar{\mathcal{P}}}(h)\ne\Delta_{H}(h)$, and is recognized to be just the Lebesgue
measure $d\mu$ on $\Re$. Then, we can construct the Hilbert space
$L^{2}(X,\nu_{s},\textbf{C})=L^{2}(\Re,d\mu)$, formed by the functions defined by $f(x)=F(s(x))$, for every
$F\in L^{2}(\bar{\mathcal{P}},H,U)$ (see the Appendix), which obviously admits a
$\bar{\mathcal{P}}$-invariant scalar product.

Solving the equation
$s(x)g=hs(xg)$ for $h=h(\theta^{+},\alpha',\beta')$, where $k=k(\theta^{1})$ and $g=g(\theta''^{a},\alpha'',\beta'')$, we can realize the induced representation
$\textrm{ind}(\bar{\mathcal{P}},H,U)$ on the separable Hilbert space $L^{2}(\Re,d\mu)$ of the
square-integrable complex functions having compact support on $\Re$ through
\begin{eqnarray}\label{eq:realization}
 [T(g)f](\theta^{1}) & = & e^{-(\alpha''/2)}\exp\Bigg[i\Bigg(
 -  \frac{\zeta^{A}\zeta_{A}\sqrt{-h}}{2B\zeta_{3}}\alpha'' + 
   \Big(\beta''  +  \frac{B}{2}\theta''^{0}\theta^{1}
  -  \frac{B}{4}\big((\theta''^{0})^{2} -  (\theta^{1}+\theta''^{1})^{2}\big)  \nonumber\\  
  & - & \frac{B}{4}
e^{-2\alpha''}(\theta''^{0}  -  \theta^{1}  -  \theta''^{1})^{2}\Big) \zeta_{3}  \Bigg)\Bigg]
f\big((\theta^{1}  +  \theta''^{1}  -  \theta''^{0}) e^{-\alpha''} \big). 
\end{eqnarray}

The corresponding representation of any $X\in\bar{\textrm{\i}}^{1}_{2}$ can be readily
calculated, yielding
\begin{eqnarray}\label{eq:repalg}
\rho(I) & = & i\zeta_{3},\qquad
\rho(J)  =  -\frac{1}{2}+i\left(-\frac{\zeta^{A}\zeta_{A}\sqrt{-h}}{2B\zeta_{3}}+
\frac{B}{2}(\theta^{1})^{2}\zeta_{3}\right)-\theta^{1}\frac{\partial}{\partial\theta^{1}},
\nonumber\\
\rho(P_{0}) & = & iB\theta^{1}\zeta_{3}-\frac{\partial}{\partial\theta^{1}},\quad\textrm{and}
\quad \rho(P_{1})  =  \frac{\partial}{\partial\theta^{1}}.
\end{eqnarray}
This representation is anti-Hermitian, and the operator identity $\rho(J)=\sqrt{-h}\left(\rho(P^{a})\rho(P_{a})+\zeta^{A}\zeta_{A}\right)/2B\rho(I)$
holds. So we can write these irrep's simply as
$T^{\zeta^{A}\zeta_{A},\zeta_{3}}(g(\theta^{a},\alpha,\beta))=\exp\left(\theta^{a}
\rho(P_{a})\right)\exp\left(\alpha\rho(J)\right)\exp\left(\beta\rho(I)
\right)$.

The irrep's
$T^{\zeta^{A}\zeta_{A},\zeta_{3}}$ are faithful, and it can be shown that, in natural units and for $B=1$, they are unitary equivalent to the irrep's of $\bar{\mathcal{P}}$ presented in Gadella
\textit{et al.}, \cite{gadella} but they are more general than the latter.
Since the quantization of the
corresponding elementary systems does not look anomalous (see Sec.~\ref{sec:5}), the irrep's
in the form $T^{\zeta^{A}\zeta_{A},\zeta_{3}}$ are the most physically interesting ones, although they do not correspond to coadjoint
orbits of $\bar{\mathcal{P}}$ which are HSM's for $\mathcal{P}$.

\subsection{Case $\zeta_{3}=0$ and $\zeta_{a}=0$}
The coadjoint orbit is the point $(0,0,\zeta_{2},0)$
in the three-dimensional hyperplane $u_{3}=0$. These coadjoint orbits are
classified by $\zeta_{2}$.
It is clear that the subalgebra
$\mathfrak{h}=\bar{\textrm{\i}}^{1}_{2}$ is subordinate to $\zeta$, since its first derived
algebra is
$[\mathfrak{h},\mathfrak{h}]=\textrm{wh}$, which is orthogonal to $\zeta$ or
$\langle \zeta,\textrm{wh}\rangle=0$. The subalgebra $\mathfrak{h}$ subordinate to $\zeta$ is also
admissible, since $\textrm{codim}\,\mathfrak{h}=0$, which is half the dimension of the coadjoint
orbit, and it satisfies Pukanszky's condition
$\zeta + \mathfrak{h}^{\bot}\subset\textrm{orb}(\zeta)$. It is not difficult to see that
there is no other admissible subalgebra subordinate to $\zeta$.

Denoting by $h(\theta^{a},\alpha,\beta)=\exp(\theta^{a}P_{a})\exp(\alpha J)
\exp(\beta I)$ the typical element of the subgroup $H$ generated by $\mathfrak{h}$, 
we can (see the Appendix) define the one-dimensional representation of $H$ by
$\chi(\theta^{a},\alpha,\beta)=U(h(\theta^{a},\alpha,\beta))=\exp(i\alpha\zeta_{2})$.
Since $H=\bar{\mathcal{P}}$ is unimodular, the space $L(\bar{\mathcal{P}},H,U)$ invariant under
right translations on
$\bar{\mathcal{P}}$ is formed by the complex functions satisfying the condition
\begin{eqnarray}\label{eq:rightinvfunc2}
F\left(h(\theta'^{a},\alpha',\beta')\cdot g(\theta^{a},\alpha,\beta)\right)  = 
\chi(\theta'^{a},\alpha',\beta')\!&F&\!\!\!(g(\theta^{a},\alpha,\beta)),
\nonumber\\
F\!\!\left(\!g\!\left(\Lambda^{a}_{\verb+ +b}(\alpha')\theta^{b}+\theta'^{a},\alpha'+\alpha,
\beta'+\beta+\frac{B}{2}\theta'^{a}\varepsilon_{ab}\Lambda^{b}_{\verb+ +c}(\alpha')\theta^{c}\right)\right)\!\!\!& = &\! 
\exp(i\alpha'\zeta_{2})F(g(\theta^{a},\alpha,\beta)). 
\end{eqnarray}
This means that the space $L(\bar{\mathcal{P}},H,U)=\textbf{C}$ is determined by the value of $F$ at
$\theta^{a}=\alpha=\beta=0$, or
$F(g(\theta^{a},\alpha,\beta))=\exp(i\alpha\zeta_{2})F(e)$,
so it is identified with the set of complex numbers.

It follows that
the Hilbert space $L^{2}(\bar{\mathcal{P}},H,U)$ is one-dimensional and it is formed by the complex
functions $F\in L(\bar{\mathcal{P}},H,U)$ for which $\Vert F\Vert^{2}<\infty$, where $\Vert F\Vert^{2}=(F,F)$ and the
$\bar{\mathcal{P}}$-invariant scalar product is given by
$(F_{1},F_{2})=\overline{F_{1}(e)}F_{2}(e)$.
Consequently (see the Appendix), we can realize 
the induced representation $\textrm{ind}(\bar{\mathcal{P}},H,U)$ on the Hilbert space
$L^{2}(\bar{\mathcal{P}},H,U)$ through
$[T(g)F](g')=\exp(i\alpha\zeta_{2})F(g')$,
where $g=g(\theta^{a},\alpha,\beta)$ and $g'=g(\theta'^{a},\alpha',\beta')$.
The corresponding representation of any $X\in\bar{\textrm{\i}}^{1}_{2}$ is given by
$\rho(I)  =  0$, $\rho(J)=i\zeta_{2}$, and $\rho(P_{a})=0$.

The representation of $\bar{\textrm{\i}}^{1}_{2}$ on the Hilbert space $\textbf{C}$
given above is clearly anti-Hermitian;
therefore, the irrep's of $\bar{\mathcal{P}}$ may be simply written
as $T^{\zeta_{2}}(g(\theta^{a},\alpha,\beta))=\exp\left(\theta^{a}
\rho(P_{a})\right)\exp\left(\alpha\rho(J)\right)\exp\left(\beta\rho(I)
\right)$,
and the operator identity $\displaystyle\rho(P^{a})\rho(P_{a})-2(B/\sqrt{-h})\rho(J)\rho(I)=
-\zeta^{A}\zeta_{A}$ holds. We note that the irrep's $T^{\zeta_{2}}$ are obviously unfaithful and
lack physical interest, although
they correspond to coadjoint orbits of $\bar{\mathcal{P}}$ which are HSM's for $\mathcal{P}$.
 
\subsection{Case $\zeta_{3}=0$ and $\zeta_{a}\ne 0$}
The
coadjoint orbit is the two-dimensional surface diffeomorphic to
$\Re^{2}$, immersed in the three-dimensional hyperplane $u_{3}=0$ and defined by the equation
\begin{equation}\label{eq:coorbfamilies}
u^{a}u_{a}=\zeta^{a}\zeta_{a}\textrm{,}
\end{equation}
which can be a hyperbolic cylinder or a half-plane translationally invariant in the direction of the
$u_{2}$ axis.
These coadjoint orbits are classified by $\zeta_{a}$ and gather into eight distinct families:
two families with $\zeta^{a}\zeta_{a}< 0$, two with $\zeta^{a}\zeta_{a}> 0$, and the other four
with $\zeta^{a}\zeta_{a}= 0$ (the $u_{2}$ axis does not belong to any family).

As we may choose any point on the
coadjoint orbit (see the Appendix), we pick $\zeta=(\zeta_{a},\zeta_{2},0)$. The subalgebra
$\mathfrak{h}=\textrm{wh}$ is subordinate to $\zeta$, since its first derived algebra is
$[\mathfrak{h},\mathfrak{h}]=(I)$, which is orthogonal to $\zeta$ or
$\langle \zeta,(I)\rangle=0$. The subalgebra $\mathfrak{h}$ subordinate to $\zeta$ is also
admissible, since $\textrm{codim}\,\mathfrak{h}=1$, which is half the dimension of the coadjoint
orbit, and it satisfies Pukanszky's condition
$\zeta + \mathfrak{h}^{\bot}\subset\textrm{orb}(\zeta)$. Since any other admissible subalgebra leads to a unitary
equivalent representation (see the Appendix), we choose $\mathfrak{h}$.

Denoting by $h(\theta^{a},\beta)=\exp(\theta^{a}P_{a})\exp(\beta I)$ the
typical element of the subgroup $H$ generated by $\mathfrak{h}$, 
we can (see the Appendix) define the one-dimensional representation of $H$ by
$\chi(\theta^{a},\beta)=U(h(\theta^{a},\beta))=\exp(i\theta^{a}\zeta_{a})$.
Due to the fact that $H=\textrm{WH}$ is unimodular, the space $L(\bar{\mathcal{P}},H,U)$ invariant under
right translations on
$\bar{\mathcal{P}}$ is formed by the complex functions satisfying the condition
\begin{eqnarray}\label{eq:rightinvfunc3}
F\left(h(\theta'^{a},\beta')\cdot g(\theta^{a},\alpha,\beta)\right)  =  
\chi(\theta'^{a}\!\!\!\!\! & , &\!\!\!\! \beta')F(g(\theta^{a},\alpha,\beta)),\nonumber\\
F\left(g\left(\theta^{a}+\theta'^{a},\alpha,
\beta'+\beta+\frac{B}{2}\theta'^{a}\varepsilon_{ab}\theta^{b}\right)\right)\!\!\!& = &\!  
\exp(i\theta'^{a}\zeta_{a})F(g(\theta^{a},\alpha,\beta)).
\end{eqnarray}
This means that the space $L(\bar{\mathcal{P}},H,U)$ is determined by the value of $F$ at
$\theta^{a}=\beta=0$.

It is not difficult to see that every element of
$\bar{\mathcal{P}}$ can be uniquely written as $g=h\cdot k$, where $h\in H$, $k\in K$, and $K$ is
the one-parameter subgroup of $\bar{\mathcal{P}}$ generated by $J\in
\bar{\textrm{\i}}^{1}_{2}$. Choosing the Borel mapping $s(x):=k$, where $x\in X=H\backslash\bar{\mathcal{P}}$ and
$x=Hg=Hhk=Hk$, we can identify the right-coset space $X$ with the subgroup
$K\subset\bar{\mathcal{P}}$, in the sense that $s(X)=K$. The bi-invariant measure on
$\bar{\mathcal{P}}$ splits into $d\mu(g)=\Delta_{H,\bar{\mathcal{P}}}(h)d\nu_{s}(x)d\nu(h)$,
where the measure on $X$ is determined by the right Haar measure on $K=\Re$,
$d\nu_{s}(x)=d\nu(s(x))$, which is $\bar{\mathcal{P}}$-invariant, since
$\Delta_{\bar{\mathcal{P}}}(h)=\Delta_{H}(h)$, and is just the Lebesgue measure
$d\mu$ on $\Re$. Then, we can construct the Hilbert space
$L^{2}(X,\nu_{s},\textbf{C})=L^{2}(\Re,d\mu)$, formed by the functions defined by
$f(x)=F(s(x))$ for every
$F\in L^{2}(\bar{\mathcal{P}},H,U)$ (see the Appendix), which obviously admits a
$\bar{\mathcal{P}}$-invariant scalar product.

Solving the equation
$s(x)g=hs(xg)$ for $h=h(\theta'^{a},\beta')$, where $k=k(\alpha)$ and $g=g(\theta''^{a},\alpha'',\beta'')$, we can realize the induced representation
$\textrm{ind}(\bar{\mathcal{P}},H,U)$ on the separable Hilbert space $L^{2}(\Re,d\mu)$ of the
square-integrable complex functions having compact support on $\Re$ through
$[T(g)f](\alpha)=\exp\left(i\Lambda(\alpha)^{a}_{\verb+ +b}\theta''^{b}\zeta_{a}\right)
f(\alpha+\alpha'')$.
The corresponding representation of any
$X\in\bar{\textrm{\i}}^{1}_{2}$ is given by
$\displaystyle \rho(I)=0$, $\displaystyle \rho(J)=\partial/\partial\alpha$, and $\displaystyle \rho(P_{a})=i
\Lambda(\alpha)^{b}_{\verb+ +a}\zeta_{b}$.

The operator identity $\displaystyle\rho(P^{a})\rho(P_{a})-2(B/\sqrt{-h})\rho(J)\rho(I)=
-\zeta^{A}\zeta_{A}$ holds and the representation of $\bar{\textrm{\i}}^{1}_{2}$ on
the Hilbert space $L^{2}(\Re,d\mu)$ is clearly anti-Hermitian, so the irrep's of $\bar{\mathcal{P}}$ may be simply written as
$T^{\zeta_{a}}(g(\theta^{a},\alpha,\beta))=\exp\left(\theta^{a}
\rho(P_{a})\right)\exp\left(\alpha\rho(J)\right)\exp\left(\beta\rho(I)
\right)$.
It can be shown that the irrep $T^{\zeta_{a}}$ is equivalent to the Wigner representation of the Poincar\'e group in (1+1) dimensions $\mathcal{P}$ (see Gadella \textit{et al.} \cite{gadella} and Ali and Antoine \cite{antoine}). We note that the irrep's $T^{\zeta_{a}}$ are unfaithful and not too interesting physically, since the quantization of the
corresponding classical elementary systems looks anomalous, although
they correspond to coadjoint orbits of $\bar{\mathcal{P}}$ which are HSM's for $\mathcal{P}$.  

\section{\label{sec:5}The Anomaly-Free Relativistic Particle in (1+1) Dimensions}

It is known that the dynamics of the relativistic particle in
a flat (1+1) dimensional space-time $M$ is described by the Lagrangian
$L_{B}=L_{0}+L_{WZ}$, where $L_{0}=-m(-h)^{-1/4}\sqrt{\dot{q}^{2}}$ and
$L_{WZ}=-(B/2)\varepsilon_{ab}\dot{q}^{a}q^{b}$. The central
charge $B$ is similar to an applied electrical force driving the particle into an uniformly
accelerated relativistic motion \cite{jackiw} and it is an additional free parameter (besides the mass $m$), fixed at the outset, that the relativistic particle theory must allow
for, due to the existence of a nontrivial two cocycle in the second cohomology group of the
Poincar\'e group in (1+1) dimensions $\mathcal{P}$. 
In fact, it was shown by Bargmann that $H^{2}_{0}(\mathcal{P},\Re)=\Re$ then, as a consequence of
the L\'evy--Leblond theorem, \cite{azcarraga} all the inequivalent Lagrangians $L_{B}$
 quasi-invariant
under $\mathcal{P}$ are classified by the central charge $B$.

However, it must be emphasized that the Lagrangian $L_{B}$ is
classically anomalous, since it is quasi-invariant under the transformations of $\mathcal{P}$, while the three conserved Noether charges
together with the identity $\{\mathcal{N}_{a},\mathcal{N}_{2},1\}$ constitute a Poisson bracket
realization of $\bar{\textrm{\i}}^{1}_{2}$, assuming $B\ne 0$
and $m\ne 0$.
Since $H^{2}_{0}(\bar{\mathcal{P}},\Re)=0$ (see Sec.~\ref{sec:1}), we can eliminate the
classical anomaly by adding a third term to $L_{B}$, depending on an extra degree of freedom
$\chi$ with dimension of action and transforming as $\chi'=\chi+\beta+(B/2)\theta^{a}\varepsilon_{ab}\Lambda^{b}_{\verb+ +c}q^{c}$ under $\bar{\mathcal{P}}$.
This addition neutralizes
the Wess--Zumino term $L_{WZ}$, causing the new Lagrangian $\bar{L}=L_{B}-\dot{\chi}$ to be
invariant under the transformations of $\bar{\mathcal{P}}$. Now, there are
four conserved Noether charges $\{\mathcal{N}_{a},\mathcal{N}_{2},\mathcal{N}_{3}\}$ associated
with the anomaly-free Lagrangian $\bar{L}$,
 which realize $\bar{\textrm{\i}}^{1}_{2}$ with the identically conserved charge
$\mathcal{N}_{3}=-1$ corresponding to the central generator realized by minus the identity.

Performing the Hamiltonian formulation of the system described by
$\bar{L}$, we learn that $\chi$ is an internal gauge degree of freedom, corresponding to the phase of the particle's wave function at the quantum level. The reduced phase-space
$\Gamma^{+}_{R}$ can be determined by observing that the constraint surface $\Gamma^{+}$ is globally diffeomorphic to $\bar{\mathcal{P}}$, such that the action of the dynamical group upon $\Gamma^{+}$ is simply
transitive and free.

It is not difficult to see that the generators of the gauge transformations
corresponding to the two primary first-class constraints $\phi_{m}$ span a subalgebra of
$\mathfrak{X}(\Gamma^{+})$ which realizes a two-dimensional Abelian subalgebra of
$\bar{\textrm{\i}}^{1}_{2}$, therefore, the reduced phase-space $\Gamma^{+}_{R}\sim\Re^{2}$
 is
diffeomorphic to the homogeneous coset space generated by the translations $P_{a}$ and can be
globally parametrized by the space--time coordinates $q^{a}$. The space
$\Gamma^{+}_{R}$ is endowed with the symplectic form
$\Omega^{+R}=d\Lambda^{+R}=(B/2)\varepsilon_{ab}dq^{a}\wedge dq^{b}$, the canonical
one form of which is given by the Wess--Zumino form $\Lambda^{+R}=(B/2)\varepsilon_{ab}q^{a}dq^{b}$.

It turns out that the symplectic manifold $(\Gamma^{+}_{R},\Omega^{+R})$ is a
Hamiltonian G-space and hence a classical relativistic elementary system. Indeed
$(\Gamma^{+}_{R},\Omega^{+R})$ is homogeneous under the
action of the dynamical group $\bar{\mathcal{P}}$, which has a Poisson action upon $\Gamma^{+}_{R}$, such that the globally Hamiltonian vector fields at $s\in\Gamma^{+}_{R}$ are given by
$\displaystyle\bar{T}^{\Gamma^{+}_{R}}_{a}(s)=(\partial/\partial q^{a})$, $\displaystyle\bar{T}^{\Gamma^{+}_{R}}_{2}
(s)=\sqrt{-h}\,\varepsilon^{a\verb+ +}_{\verb+ +b}q^{b}(\partial/\partial q^{a})$, and
$\displaystyle\bar{T}^{\Gamma^{+}_{R}}_{3}(s)=0$. The comoments are given by
$u^{+R}_{a}(s)=Bq^{b}\varepsilon_{ba}$, $u^{+R}_{2}(s)=(m^{2}/2B)+[B/(2\sqrt{-h})]q_{a}q^{a}$, and $u^{+R}_{3}(s)=-1$. Note that they are not uniquely determined, since $u^{+R}_{2}$ is defined up to an additive
constant, consistently with $H^{1}_{0}(\bar{\mathcal{P}},\Re)=\Re$.

The identities
$u^{+R}_{A}u^{+RA}(s)=m^{2}/\sqrt{-h}$ and $u^{+R}_{3}(s)=-1$ hold, so $u^{+R}_{2}(s)$ is functionally dependent on the $u^{+R}_{a}(s)$, which are regarded as the fundamental dynamical variables, and
using the fact that the comoments constitute a Poisson bracket realization of $\bar{\textrm{\i}}^{1}_{2}$, it is not difficult to see that
$\{ q^{a},q^{b}\}=[\varepsilon^{ab}(\sqrt{-h})^{2}]/B$.
The value of the momentum mapping 
$\displaystyle\mu^{+}_{R}(s)=([u^{+R}_{A}(s)]/\hbar)\bar{\omega}^{A}$ at the origin $s_{0}=(0,0)$ in $\Gamma^{+}_{R}$ shall be denoted by
$\zeta=\mu^{+}_{R}(s_{0})=(0,0,m^{2}/(2B\hbar),-1/\hbar)$, which satisfies
\begin{equation}\label{eq:ressalta1}
\zeta^{A}\zeta_{A}=\frac{m^{2}}{\sqrt{-h}\hbar^{2}}\quad\textrm{and}\quad
\zeta_{3}=-\frac{1}{\hbar}.
\end{equation}
The second identity in Eq.~(\ref{eq:ressalta1}) follows from the value of
$u^{+R}_{3}(s)$ and the definition of the momentum mapping,
so the quantization of $(\Gamma^{+}_{R},\Omega^{+R})$
satisfies Dirac's quantum
condition (this will be shown later).

Moreover, a straightforward calculation shows that \cite{ricardo}
the momentum mapping $\mu^{+}_{R} :\Gamma^{+}_{R}\to orb(\zeta)$
is a symplectomorphism between
the elementary system $(\Gamma^{+}_{R},\Omega^{+R})$ and the coadjoint orbit $(orb(\zeta),b)$
through $\zeta\in\bar{\textrm{\i}}^{1^{*}}_{2}$, with $\displaystyle\mu^{+*}_{R}b  = 
(\Omega^{+R}/\hbar)$.
It follows that Eq.~(\ref{eq:ressalta1}) provides a physical interpretation for the parameters labeling
the irrep $T^{\zeta^{A}\zeta_{A},\zeta_{3}}$ of
$\bar{\mathcal{P}}$, which corresponds to the relativistic elementary system $(\Gamma^{+}_{R},\Omega^{+R})$. 

\subsection{Quantization of the anomaly-free relativistic particle in (1+1) dimensions}
Before we address the quantum dynamics of the relativistic particle, though, let us
clear up the quantization of the system at the kinematical level. Let $\varphi(\bar{T}_{A}):=i
\rho(\bar{T}_{A})$ be the Hermitian
representation of $\bar{\textrm{\i}}^{1}_{2}$ on the Hilbert space $L^{2}(\Re,dx)$ defined
from Eq.~(\ref{eq:repalg}), for $\zeta=(0,0,-[(\zeta^{A}\zeta_{A}\sqrt{-h})/2B\zeta_{3}],\zeta_{3})$ satisfying Eq.~(\ref{eq:ressalta1}), and $\mathfrak{j}\cong\bar{\textrm{\i}}^{1}_{2}$ be the finite-dimensional Lie subalgebra of $C^{\infty}(\Gamma^{+}_{R})$ spanned by
the comoments $\{u^{+R}_{A}\}$. In addition, let $\lambda:\bar{\textrm{\i}}^{1}_{2}
\mapsto C^{\infty}(\Gamma^{+}_{R})$ be the lift of the mapping
$\sigma:\bar{\textrm{\i}}^{1}_{2}\mapsto \mathcal{A}(\Gamma^{+}_{R})$ induced by the left
action of $\bar{\mathcal{P}}$ on $\Gamma^{+}_{R}$, where
$\mathcal{A}(\Gamma^{+}_{R})$ denotes the set of all the globally Hamiltonian vector fields on $\Gamma^{+}_{R}$. The mapping $\lambda(\bar{T}_{A})=u^{+R}_{A}$ is a Lie algebra homomorphism and it is well-defined, since $\Gamma^{+}_{R}$ is simply connected and $H^{2}_{0}(\bar{\textrm{\i}}^{1}_{2},\Re)=0$.

Then, $\textrm{orb}(\zeta)$ determines the linear map
$\displaystyle\mathcal{Q}:=(1/\zeta_{3})\varphi\circ\lambda^{-1}$
from $\mathfrak{j}$ onto the linear
space $\textrm{Op}(D)=\textrm{span}\{\mathcal{Q}(u^{+R}_{A})\}$ of (in general) unbounded
Hermitian (or symmetric) operators preserving a fixed dense domain $D$ in $L^{2}(\Re,dx)$,
which satisfies
$\mathcal{Q}(\{u^{+R}_{A},u^{+R}_{B}\})  =  -i\zeta_{3}[\mathcal{Q}(u^{+R}_{A}),\mathcal{Q}(u^{+R}_{B})]$ and $\mathcal{Q}(u^{+R}_{3})  =  -\mathbf{1}$.
For the domain $D$, we can take the Schwartz space $\mathbf{\mathcal{S}}(\Re,\textrm{\textbf{C}})\subset L^{2}(\Re,dx)$ of rapidly decreasing smooth complex-valued functions, for instance.
     
Recalling that $u^{+R}_{3}=-1$, we can see that Dirac's quantum condition is satisfied if and
only if $\zeta_{3}=-(1/\hbar)$\label{sec:dirac}, consistently with
Eq.~(\ref{eq:ressalta1}). Furthermore, assuming that $D$ is a domain of essential
self-adjointness for $\textrm{Op}(D)$, we can see that the linear map $\mathcal{Q}$ is
actually a prequantization of $\mathfrak{j}$ in the sense of Gotay, \cite{go3} since the
globally Hamiltonian vector fields $\bar{T}^{\Gamma^{+}_{R}}_{A}$ are complete. It is not
difficult to see that $\mathbf{\mathcal{S}}(\Re,\textrm{\textbf{C}})$ is a domain of
essential self-adjointness for the representation of $\mathfrak{j}$
given by $\textrm{Op}(D)$.

In order to determine the maximal Lie subalgebra $\mathcal{O}$ of
$C^{\infty}(\Gamma^{+}_{R})$ that can consistently be quantized, we will tie to the approach
that aims at providing a quantization
of the pair $(\mathcal{O},\mathfrak{b})$, i.e., a prequantization of $\mathcal{O}$ which (among
other things) irreducibly represents a suitably chosen basic algebra of observables
$\mathfrak{b}\subset C^{\infty}(\Gamma^{+}_{R})$.\cite{go3}
It turns out that the suitable basic
algebra is $\mathfrak{b}=\textrm{wh}=\textrm{span}\{u^{+R}_{0},
u^{+R}_{1},u^{+R}_{3}\}$, since the restriction of $\mathcal{Q}$ to
$\textrm{wh}\subset\mathfrak{j}=\textrm{span}\{u^{+R}_{A}\}$ provides actually a quantization of the pair $(\mathfrak{b},\mathfrak{b})$, which is equivalent to the usual
Schr\"odinger quantization of a one-dimensional nonrelativistic free particle.

In fact, in the coordinates of $\Gamma^{+}_{R}$ defined by
$q:=-(u^{+R}_{0}+u^{+R}_{1})/B=-q^{1}+q^{0}$ and $p:=u^{+R}_{1}=-Bq^{0}$,
the expression of the associated quantization map $\mathcal{Q}$ is exactly given by the Schr\"odinger representation of wh in the position
representation $\{\vert x\rangle\}$;
$\hat{q}:=\mathcal{Q}(q)=x$, $\hat{p}:=\mathcal{Q}(p)=-i\hbar(\partial/\partial x)$, and
$\hat{1}:=\mathcal{Q}(1)=\mathbf{1}$
on the domain $D$, such as $D=\mathbf{\mathcal{S}}(\Re,\textrm{\textbf{C}})\subset
L^{2}(\Re,dx)$. The standard canonical quantization is well-defined; however,
there is no full quantization of $(C^{\infty}(\Gamma^{+}_{R}),\mathfrak{b})$ in which a
Von Neumann rule is compatible with the Schr\"odinger quantization.

Indeed, due to the strong Groenewold--Van Hove no-go theorem,\cite{go3} there is no
quantization of $(P,\textrm{wh})$ on $\Re^{2}\sim \Gamma^{+}_{R}$, where
$P$ denotes the polynomial subalgebra of $C^{\infty}(\Gamma^{+}_{R})$
generated by $\mathfrak{b}=\textrm{wh}$.
It turns out
that the only two distinct isomorphism classes of maximal Lie subalgebras of $P$ which contain
$\textrm{wh}$ are those represented by $P^{2}$ and by the set of polynomials
$S=\{f(q)p+g(q)\}$,
where $P^{2}$ denotes the subspace of polynomials of degree at most 2, and $f,g$ are polynomials.\cite{go3}

The quantization of the pair $(P^{2},\textrm{wh})$ is provided by the well-known extended metaplectic quantization. On the other hand,
the only classical observable in this paper that will require it
is the comoment $u^{+R}_{2}$, which is in $P^{2}\subset P$ but not in $\textrm{wh}$. For all the other observables that we will consider, such as
position, momentum, potential energy,
relativistic energy, or the Hamiltonian, the Schr\"odinger quantization will be enough. In particular, we will not consider any observable in $S$.

In addition, a straightforward calculation shows that
the extended metaplectic quantization $\mathcal{Q}$ of
$(P^{2},\textrm{wh})$ is covariant with respect to $\bar{\mathcal{P}}$ in the sense that, for every
$f(q,p)$ in $P^{2}\subset P\subset C^{\infty}(\Gamma^{+}_{R})$ and $g=g(\theta^{a},\alpha,\beta)\in\bar{\mathcal{P}}$, we have
$\mathcal{Q}(f(q',p'))=T^{\zeta^{A}\zeta_{A},\zeta_{3}}(g^{-1})\mathcal{Q}(f(q,p))T^{\zeta^{A}\zeta_{A},\zeta_{3}}(g)$,
where $(q',p')=l_{g}(q,p)$ is the left action on $\Gamma^{+}_{R}$ generated by the globally
Hamiltonian vector fields $\bar{T}^{\Gamma^{+}_{R}}_{A}$, and $\zeta$ satisfies Eq.~(\ref{eq:ressalta1}).

\subsection{Quantum dynamics of the anomaly-free relativistic particle in (1+1) dimensions}

As far as the quantum dynamics of the system is concerned, we remark that the total energy of
the particle depends explicitly on time, so it does not provide a suitable Hamiltonian.  
For this reason, we
turn to consider the dynamics from the point of view of the reduced phase-space $\Gamma^{+}_{R}$.
The central charge determines the symplectic form
$\Omega^{+R}=-B\,\textrm{vol}$, which is proportional to the volume two form of space--time and can be expressed in the coordinates $(q,p)$ of $\Gamma^{+}_{R}$ by $\Omega^{+R}=-dp\wedge dq$, with the Wess--Zumino form given by minus the Liouville form $\Lambda^{+R}=-pdq$.

Up to gauge equivalence, the dynamics on $\Gamma^{+}_{R}$ is specified by 
$q^{0}(\tau)=\tau$,
\begin{displaymath}
q^{1}(\tau)  =  q^{1}(\tau_{0})-\sqrt{m^{2}+
\tilde{p}(\tau_{0})^{2}}/B+\sqrt{m^{2}+\tilde{p}(\tau)^{2}}/B,
\end{displaymath}
and $\tilde{p}(\tau)=
\tilde{p}(\tau_{0})+B(\tau-\tau_{0})$, for a given $\tilde{p}(\tau_{0})$,with $\tau_{0}\in\Re$. It follows
that the proper time is given by $t'=(m/B)\textrm{arsinh}[\tilde{p}(\tau)/m]$, and
$\tilde{p}(\tau)$ is the kinematical momentum, since $\displaystyle\tilde{p}(\tau)=\gamma(\tau)
m(dq^{1}/dt)(\tau)$.

Note
that the equations for $q^{a}(\tau)$ are regarded as Hamilton equations, while that for
$\tilde{p}(\tau)$ is an identity. Moreover, retaining the space--time meaning of
the reduced phase-space, the world line $W$ of the particle is also a Hamiltonian flow in the
symplectic manifold $\Gamma^{+}_{R}$. 
Calculating the globally Hamiltonian vector field corresponding to this flow, 
$X_{H}(\tau)=\bar{T}^{\Gamma^{+}_{R}}_{0}(\tau)+[\tilde{p}(\tau)/\sqrt{m^{2}+
\tilde{p}(\tau)^{2}}]\bar{T}^{\Gamma^{+}_{R}}_{1}(\tau)$, and applying the antihomomorphism of
Lie algebras $\lambda\circ\sigma^{-1}(\bar{T}^{\Gamma^{+}_{R}}_{a})=
u^{+R}_{a}$, we get the Hamiltonian $H(q,p,\tau)$.

The corresponding Hamiltonian operator splits into two parts
$\hat{H}(\hat{q},\hat{p},\tau)=\hat{H}_{0}(\hat{q},\hat{p})
+\hat{V}(\hat{p},\tau)$,
where $\hat{H}_{0}(\hat{q},\hat{p})=-B\hat{q}-\hat{p}$ and
$\hat{V}(\hat{p},\tau)=[\tilde{p}(\tau)/\sqrt{m^{2}+\tilde{p}(\tau)^{2}}]\hat{p}$. Solving
the eigenvalue problem $\hat{H}_{0}\vert E\rangle=E\vert E\rangle$, we discover that
$\hat{H}_{0}$ has continuous spectrum with the normalized eigenfunctions given by $\displaystyle
\langle x
\vert E\rangle=(1/\sqrt{2\pi\hbar})\exp[-(i/\hbar)(Ex+(B/2)x^{2})]$, so $\langle E'\vert E\rangle=\delta(E'-E)$.

Note that classically
$H_{0}=u^{+R}_{0}=Bq^{1}=-2\mathcal{E}_{\mathrm{pot}}(q^{1})$, so  
$\hat{H}_{0}(\hat{q},\hat{p})=-2\hat{\mathcal{E}}_{\mathrm{pot}}(\hat{q},\hat{p})$ has the meaning of a potential energy operator. Besides
this fact, the total energy operator $\hat{\mathcal{H}}(\hat{q},\hat{p},\tau)=\mathcal{E}(\tau)
-\frac{1}{2}\hat{H}_{0}(\hat{q},\hat{p})$,where $\mathcal{E}(\tau):=\sqrt{m^{2}+\tilde{p}(\tau)^{2}}$ is the relativistic energy of the particle, satisfies $[\hat{\mathcal{H}},\hat{H}_{0}]=0$;
therefore
the eigenvectors of $\hat{H}_{0}$ are simultaneously total energy eigenstates. Then, the eigenvalues
of the total energy operator are related with those of $\hat{H}_{0}$ through
$\hat{\mathcal{H}}(\tau)\vert E\rangle=E_{T}(\tau)\vert E\rangle$, where
$E_{T}(\tau)=\mathcal{E}(\tau)-E/2$.

In terms of the base kets
$\{\vert E\rangle\}$, the state ket of the system is given at $\tau=\tau_{0}$ by $\vert\alpha
\rangle=\int^{\scriptscriptstyle +\infty}_{\scriptscriptstyle -\infty}dE\, c_{E}(\tau_{0})\vert E\rangle$, where $c_{E}(\tau_{0})$ is some
known complex function of $E$ satisfying $\int^{\scriptscriptstyle +\infty}_{\scriptscriptstyle -\infty}dE\vert c_{E}(\tau_{0})\vert^{2}=1$.
Then, for $\tau>\tau_{0}$, the state ket will be $\vert\alpha,\tau_{0};\tau
\rangle=\int^{\scriptscriptstyle +\infty}_{\scriptscriptstyle -\infty}dE\, c_{E}(\tau)e^{(iE/\hbar)(\tau-\tau_{0})}\vert E\rangle$,
where the $c_{E}(\tau)$'s satisfy the coupled differential equations
\begin{displaymath}
-i\hbar\frac{dc_{E}}{d\tau}
(\tau)=\int^{+\infty}_{-\infty}dE'\langle E\vert \hat{V}\vert E'\rangle e^{[-i(E-E')/\hbar]
(\tau-\tau_{0})}c_{E'}(\tau). 
\end{displaymath}

Solving the resulting linear homogeneous partial differential equations, we get
\begin{displaymath}
c_{E}(\tau)=\exp\!\Bigg[\!\frac{i(
(\!E\!-\!\Delta\mathcal{E})^{2}\! -\!
E^{2})}{2B\hbar}\!\Bigg]\!\cdot\!\exp\!\Bigg[\!-\frac{i}{\hbar}\Big(\!-\frac{\tilde{p}(\tau_{0})\Delta\mathcal{E}}{2B}-\frac{m\Delta t'}{2}+\frac{\mathcal{E}\Delta\tau}{2}\Big)\!\Bigg]\!\cdot\!
c_{E-\Delta\mathcal{E}}(\tau_{0}),
\end{displaymath}
where $\Delta\mathcal{E}=\sqrt{m^{2}\!+\!\tilde{p}(\tau)^{2}}\! - \! \sqrt{m^{2}\!  +\!  \tilde{p}(\tau_{0})^{2}}$, $\Delta t'= (m/B)(\textrm{arsinh}[\tilde{p}(\tau)/m]-\textrm{arsinh}[\tilde{p}(\tau_{0})/m])$, and $\Delta\tau = \tau-\tau_{0}$.

Suppose now that the system is initially prepared in an energy
eigenstate $\displaystyle\vert\alpha\rangle=\vert E\rangle$, then at a later time $\tau>\tau_{0}$ the state
will be given by
\begin{eqnarray}\label{eq:latertau}
\vert\alpha,\!\tau_{0}\,;\tau\rangle\! & = &\!  \exp\!\Bigg[\!\frac{i(E^{2}
-(\!E\!+\!\Delta\mathcal{E}\,)^{2})}{2B\hbar}\!\Bigg]\!\cdot\!
 \exp\!\Bigg[\!\frac{i(\!E\!+\!\Delta\mathcal{E}\,)\Delta\tau}{\hbar}\Bigg]\!\cdot{}\nonumber\\
& & {}\exp\!\Bigg[\!-\frac{i}{\hbar}\Big(\!-\frac{\tilde{p}(\tau_{0})\Delta\mathcal{E}}{2B}-\frac{m\Delta t'}{2}+\frac{\mathcal{E}\Delta\tau}{2}\Big)\!\Bigg]\!\cdot\!\vert E\! +\! \Delta\mathcal{E}\rangle.
\end{eqnarray}

The probability as a function of time for the particle to be found in the state $\vert E'\rangle$
is given by
$\vert\langle E'\vert\alpha,\tau_{0};\tau\rangle\vert^{2}/\langle
\alpha,\tau_{0};\tau\vert\alpha,\tau_{0};\tau\rangle dE'=\delta(E'-E -
\sqrt{m^{2}+\tilde{p}(\tau)^{2}} + \sqrt{m^{2}+\tilde{p}(\tau_{0})^{2}})dE'$,
which equals 1 if
$E'= E+\sqrt{m^{2}+\tilde{p}(\tau)^{2}}- \sqrt{m^{2}+\tilde{p}(\tau_{0})^{2}}$ or zero
otherwise. From Eq.~(\ref{eq:latertau}), we note that the states $\vert E\rangle$ are not stationary
although they are total energy eigenstates, since the $\tau$-dependent part of the Hamiltonian
$\hat{V}(\hat{p},\tau)$ causes transitions to eigenstates
$\vert E + \sqrt{m^{2}+\tilde{p}(\tau)^{2}} -
\sqrt{m^{2}+\tilde{p}(\tau_{0})^{2}}\rangle$ of different energy.

In fact, the
expectation value of the total energy operator, for instance,  
\begin{displaymath}
\langle\hat{\mathcal{H}}\rangle(\tau)=
\frac{\langle\alpha,\tau_{0};\tau\vert\hat{\mathcal{H}}\vert\alpha,\tau_{0};\tau\rangle}{\langle
\alpha,\tau_{0};\tau\vert\alpha,\tau_{0};\tau\rangle}=\frac{\mathcal{E}(\tau)}{2}+\frac{\mathcal{E}(\tau_{0})}{2}-\frac{E}{2},
\end{displaymath}
is $\tau$-dependent. It is not difficult to see that the function
$\langle\hat{\mathcal{H}}\rangle(\tau)$ attains to a minimum at $\tau=\tau_{0}-
[\tilde{p}(\tau_{0})/B]$, when its value is $\langle\hat{\mathcal{H}}\rangle(\tau_{0}-
[\tilde{p}(\tau_{0})/B])=m/2+[\sqrt{m^{2}+\tilde{p}(\tau_{0})^{2}}/2]-(E/2)$, which only happens after $\tau_{0}$ if $\tilde{p}(\tau_{0})$ satisfies the condition
$\textrm{sign}(B)\tilde{p}(\tau_{0})<0$; otherwise, $\langle\hat{\mathcal{H}}\rangle(\tau)$ is a
monotonically increasing function of $\tau>\tau_{0}$. For this reason, the presented quantum
states are stable, even though there is no true ground state.

\section{\label{sec:6}Discussion}

We showed that the extended Poincar\'e group in (1+1)
dimensions $\bar{\mathcal{P}}$ is a connected solvable
exponential Lie group, with $H^{2}_{0}(\bar{\mathcal{P}},\Re)=0$ and
$H^{1}_{0}(\bar{\mathcal{P}},\Re)=\Re$ (see Sec.~\ref{sec:1}). These facts
were important to apply the Kirillov theorem to perform a
classification of the two-dimensional relativistic elementary systems
and to work out explicitly all the irrep's of $\bar{\mathcal{P}}$ by the orbit method (see Sec.~\ref{sec:4}).
The particular class of irrep's $T^{\zeta^{A}\zeta_{A},\zeta_{3}}$
with $\zeta$ satisfying Eq.~(\ref{eq:ressalta1}) turned out to be
connected to a covariant maximal polynomial quantization of the anomaly-free relativistic particle in (1+1)
dimensions, which provided a quantum-mechanical interpretation
for the construction in this most physically interesting case (see Sec.~\ref{sec:5}). 

We remark that the Bohr--Wilson--Sommerfeld condition\cite{woodhouse} is trivially satisfied by the anomaly-free relativistic particle in (1+1)
dimensions, and it does not yield the quantization of any observable
quantity, which is consistent with the fact that the system is not conservative
and the world lines are open. It is worth mentioning that
$\bar{\mathcal{P}}$ is related to
the one-dimensional oscillator group Os(1) by the Weyl unitary trick. However, the group Os(1)
is not solvable exponential, and the orbit method gives all its irrep's only
through holomorphic induction.\cite{streater}

It is remarkable that the extended metaplectic quantization of the anomaly-free relativistic particle is covariant with respect to $\bar{\mathcal{P}}$, inasmuch as 
a covariant Stratonovich--Weyl kernel for the corresponding coadjoint orbits has
not been found yet. \cite{gadella} However, this difficulty is not directly
related to the fact that there is an obstruction to fully
quantizing the system, since there are symplectic manifolds such as
$\Re^{2}$ or $S^{2}$ for which the problem of the generalized Weyl--Wigner--Moyal
quantization has successfully been  solved, \cite{gadella} although obstructions have
been found. \cite{go3}
 
The group $\bar{\mathcal{P}}$ enjoys several properties in common with other Lie groups of low dimension such as the Weyl--Heisenberg group 
WH, the Euclidean group E(2), and the affine group $\textrm{Aff}_{+}(1,\Re)$, which found
applications in fields such as electronics, signal processing, and quantum optics.
As all these groups have
square-integrable representations,
in a subsequent publication it would be interesting to test whether
the irrep's $T^{\zeta^{A}\zeta_{A},\zeta_{3}}$ of $\bar{\mathcal{P}}$ 
are square integrable with respect to $\textrm{orb}(\zeta)$. This fact would allow us to work out the associated generalized coherent states, 
generalized wavelet transforms, and generalized Wigner functions,\cite{ali} which
would surely be an invaluable mathematical tool in the context of the phase-space
formulation of the quantum anomaly-free relativistic particle in (1+1) dimensions.

\begin{acknowledgments}

R. O. de M. would like to thank R. F. Streater for reading
the preprint, R. Aldrovandi for the discussions about the orbit method, and N. Grishkov for
helping us to calculate the second cohomology group of the extended Poincar\'e algebra.
This
research was partially supported by CNPq.  V. O. R. acknowledges support from PRONEX under
Contract CNPq 66.2002/1998-99.

\end{acknowledgments}

\appendix

\section*{The Method of Orbits}

Before considering
the Bernat--Pukanszky theory of exponential groups,\cite{bernat, puk} let us briefly review
the standard procedure to form a unitary induced representation, since the theory of induced
representations developed by Mackey \cite{mackey1, mackey2} plays
an essential role in the method of orbits.  

Let $H$ be a closed subgroup of a locally compact topological group with a countable basis $G$
and $U$ a one-dimensional unitary representation of $H$ on the complex numbers
\textbf{C}. We introduce the space $L(G,H,U)$ of complex-valued measurable functions $F$ on $G$
that satisfy the condition $F(hg)=\Delta_{H,G}(h)^{-1/2}U(h)F(g)$,
where $\Delta_{H,G}(h)=\Delta_{H}(h)/ \Delta_{G}(h)$, $h\in H$, $g\in G$, and
$g\mapsto\Delta_{G}(g)$ is the modulus of the group $G$.   

The group $G$ can be identified with $H\times X$, where $X$ is the right $G$-space $X=H\backslash G$, since every element of $g\in G$ can be written uniquely in the form $g=hs(x)$ with $x\in X$. Under this identification, the right Haar measure on $G$ splits into the product of a quasi-invariant measure $\nu_{s}$ on $X$, depending upon the choice of a Borel mapping $s$ of $X$ into $G$
having the property that $s(Hg)\in Hg$, by the right Haar measure on $H$;       
$d\nu(g)=\Delta_{H,G}(h)d\nu_{s}(x)d\nu(h)$. The measure $\nu_{s}$ on $X$ is $G$-invariant  if
and only if $\Delta_{G}(h)=\Delta_{H}(h)$.

The space $L(G,H,U)$ is clearly invariant under right translations on $G$.
Let $L^{2}(G,H,U)$ denote the Hilbert space generated by the square-integrable functions
$F$ in $L(G,H,U)$; then, we call the unitary representation $T$ acting by right translations upon the Hilbert space
$L^{2}(G,H,U)$, according to
$[T(g)F](g')=F(g'g)$,
the representation induced in the sense of Mackey by the representation $U$ and
we will denote it by $\textrm{ind}(G,H,U)$.

It is not difficult to see that
there is an isomorphism $F\mapsto f$ of the Hilbert space
$L^{2}(G,H,U)$ onto the Hilbert space $L^{2}(X,\nu_{s},\textbf{C})$, generated by the
square-integrable complex functions having compact support on $X$ with respect to the measure
$\nu_{s}$, which associates a function $f\in L^{2}(X,\nu_{s},\textbf{C})$ defined by
$f(x)=F(s(x))$ with every  $F\in L^{2}(G,H,U)$.
Under this isomorphism, the induced representations in the sense of Mackey can be realized on the
Hilbert space $L^{2}(X,\nu_{s},\textbf{C})$ through
$[T(g)f](x)=\Delta_{H,G}(h)^{-1/2}U(h)f(xg)$,
where the element $h\in H$ is defined from the relation $s(x)g=hs(xg)$.

Now, we can sketch the orbit method. Let $G$ be an exponential group, $\mathfrak{g}$ its
real exponential Lie algebra, and
$\mathfrak{g}^{*}$ its dual.
We say that a subalgebra $\mathfrak{h}\subset \mathfrak{g}$ is
subordinate to $\zeta\in\mathfrak{g}^{*}$ if its first derived algebra is
orthogonal to $\zeta$, or $\langle \zeta,[\mathfrak{h},\mathfrak{h}]\rangle=0$. Denoting by
$H\subset G$ the subgroup corresponding to the subalgebra $\mathfrak{h}$ subordinate to
$\zeta\in\mathfrak{g}^{*}$, we define the unitary one-dimensional representation of $H$ by
$U(\exp X)=\exp (i\langle \zeta, X\rangle)$, which is  
related to the character $\chi$ of $H$ simply by $\chi(\exp X)=U(\exp X)$, where
$X\in\mathfrak{h}$.

Then a unitary induced representation $\textrm{ind}(G,H,U)$ of $G$ is irreducible
if and only if the subalgebra
$\mathfrak{h}$ subordinate to $\zeta\in\mathfrak{g}^{*}$ is admissible, i.e., if its dimension
is maximal in the family of all subalgebras subordinate to $\zeta$ and if it satisfies
Pukanszky's condition. \cite{puk} The maximality condition is equivalent to $\textrm{dim}\,\mathfrak{h}=\textrm{dim}\,\mathfrak{g}-\frac{1}{2}\textrm{dim}\,\textrm{orb}(\zeta)$, and Pukanszky's condition requires that the
linear variety $\zeta + H^{\bot}$ is contained in $\textrm{orb}(\zeta)$, where $H^{\bot}$
denotes the orthogonal complement of $H$ in $\mathfrak{g}^{*}$. Bernat \cite{bernat} showed that
the first condition implies the second one if $\mathfrak{g}$ is quasinilpotent (i.e., all
the real eigenvalues of $\textrm{ad}(X)$ are zero, for all $X\in\mathfrak{g}$), otherwise the two conditions are independent. In particular, every nilpotent group is quasinilpotent.  

It can be shown \cite{puk2} that, for any given $\zeta$, there is a subordinate subalgebra
$\mathfrak{h}$ satisfying the two conditions above.  
Moreover, if $\mathfrak{h}_{1}$ and $\mathfrak{h}_{2}$ are, respectively, maximal dimension
subalgebras subordinate to $\zeta_{1}$ and $\zeta_{2}$, further obeying Pukanszky's condition, then
$\textrm{ind}(G,H_{1},U_{1})=\textrm{ind}(G,H_{2},U_{2})$ if and
only if $\zeta_{1}$ and $\zeta_{2}$ belong to the same coadjoint orbit, the
equal sign indicating unitary equivalence. Reciprocally, any irrep
of $G$ is representable in the form $\textrm{ind}(G,H,U)$ by specifying $\mathfrak{h}$ and $\zeta$
appropriately, thus establishing a canonical bijection between the space $\mathcal{O}(G)$ of coadjoint orbits and the unitary dual $\widehat{G}$ of any solvable exponential Lie group. It is worth mentioning that every
coadjoint orbit of the connected and
simply connected solvable type I Lie groups (and, in
particular, of the exponential groups) is integral (i.e., satisfies the integrality condition).

\end{document}